\documentclass[
    final             % use final for the camera ready runs
    ,numberedheadings % uncomment this option for numbered sections
  ]
  {aipproc}

\layoutstyle{8x11single}

\def\fm{\hbox{$.\!\!^{\rm m}$}}
\def\degr{\hbox{$^\circ$}}
\def\sun{\hbox{$\odot$}}
\def\kms{\mbox{km\,s$^{-1}$\/}}

\begin{document}

\title{Interpretation of the Line Spectrum of Classical Symbiotic Stars in the Scenario for their Prototype Z~And\thanks{Based on observations collected at the National Astronomical Observatory Rozhen, Bulgaria}}


\author{N.~A.~Tomov}{
  address={Institute of Astronomy and National Astronomical Observatory, Bulgarian Academy of Science,\\ PO Box 136, BG-4700 Smolyan, Bulgaria\\}
}

\author{D.~V.~Bisikalo}{
  address={Institute of Astronomy, Russian Academy of Science, 48 Pyatnitskaya Street, RU-119017 Moscow, Russia}
}

\author{M.~T.~Tomova}{
  address={Institute of Astronomy and National Astronomical Observatory, Bulgarian Academy of Science,\\ PO Box 136, BG-4700 Smolyan, Bulgaria\\}
}

\author{E.~Yu.~Kil'pio}{
  address={Institute of Astronomy, Russian Academy of Science, 48 Pyatnitskaya Street, RU-119017 Moscow, Russia}
}

\begin{abstract}
Results of the study of the symbiotic binary Z~And during its recent active phase 2000--2010 when it experienced a series of six optical outbursts are presented. High-resolution spectra obtained during the first and fourth outburst, which was the strongest one, have been analyzed. These data are compared with results of theoretical computations. The comparison provides information about the behavior of the system during the entire active phase rather than during an individual outburst. In particular it was found fundamental difference between the first outburst which opened the active phase and the recurrent outbursts---namely, the presence of bipolar collimated optical outflow during some of the recurrent outbursts. A scenario that can explain all the spectroscopic phenomena observed during this active phase as well as previous active phases of Z~And is proposed. The possibility to use this scenario for explanation of the line spectrum of other classical symbiotic stars during their active phases is motivated.
\end{abstract}

\maketitle

%%%%%%%%%%%%%%%%%%%%%%%%%%%%%%%%%%%%%%%%%%%%
%% MAINMATTER
%%%%%%%%%%%%%%%%%%%%%%%%%%%%%%%%%%%%%%%%%%%%

\section{Introduction}
\label{sec:intro}

Symbiotic stars are interpreted as interacting binaries consisting of a cool visual primary and a hot compact secondary component accreting matter from the atmosphere of its companion. Their spectral variability is determined from the orbital motion and the outburst events of the hot component. These events are often accompanied by intensive loss of mass in the form of optically thick shells, stellar wind outflow and bipolar collimated jets. The stellar wind is possible to be bipolar and highly collimated too. The ejected material forms complicated flow structure in the space between the components, whose elements determine the observed properties of the star. The interacting binary Z And is considered as a prototype of the classical symbiotic stars. It consists of a normal cool giant of spectral type M4.5 \citep{MS}, a hot compact component with a temperature of about $1.5\times10^5$ K \citep{Sok06} and an extended circumbinary nebula formed by the winds of the stellar components and partly photoionized by the compact object.

Z And has undergone several active phases consisting of repeated optical brightenings with amplitudes up to $2$--$3$ mag and characterized by intensive loss of mass \citep{SS,Boyarchuk,FC95,Bis06,Sk06,Sok06,TTB08,Sk09,TTB10}. The last active phase of Z And began at the end of August 2000 \citep{Sk00} and continued up to that time including six optical brightenings. The maxima of the light during the active phase were in December 2000 ($ V\sim 8$\fm8), November 2002 ($V \sim 9$\fm8), September 2004 ($V \sim 9$\fm1), July 2006 ($V \sim 8$\fm6), January 2008 ($V\sim 9$\fm5) and January 2010 ($V \sim 8$\fm4) \citep{Sk09} (we also used data of the American Association of Variable Star Observers, AAVSO, and via private communication from A.~Skopal). High resolution optical data indicating highly different physical conditions (velocity, density, temperature) in the line emission regions were obtained during this active phase \citep{Sk06,Sok06,TTB07,TTB08,Sk09,TTB10}.

The current study is based on observations acquired during the first and fourth outbursts. The main features of the flow structure in the system during the active phase indicated by our data can be summarized as follows:
\begin{enumerate}
	\item We suppose that an accretion disk is present in the system. Only in this way we can explain the presence of the \mbox{He\,{\sc ii}} $\lambda$4686 line in the spectrum and its behavior during the first outburst \citep{TTB08}.
	\item Stellar wind was observed in the system during all the outbursts. Two wind components with different velocity regimes were observed simultaneously---wind propagating at a moderate velocity, $0$--$100$ \kms \citep{Sok06,Sk06,TTB08}, and wind with a high velocity, in excess of $500$ \kms \citep{Sk06,TTB08}.
	\item Bipolar collimated outflow appeared during the fourth outburst. In the beginning of July 2006 the H$_\alpha$ line had an absorption component shifted by $-1400$ \kms\, from its center. It went into emission and later, in July--December 2006, the spectrum contained additional emission components on either side of the central peak corresponding to velocities of $1200$--$1500$ \kms\, \citep{BL,TTB07,Sk09}. The H$_\beta$ line had the same features \citep{TTB07,Sk09}. Both the emission and absorption high-velocity components are assumed to be formed in a bipolar collimated outflow \citep{TTB07}.
\end{enumerate}

The main aim of the current study is to suggest a model for the flow structure in the system to explain all of these features. We believe that it will be helpful in explaining not only the data acquired during this active phase, but also all spectroscopic data taken during earlier phases, those following 1939, 1960 and 1984 and possibly the activity of other classical symbiotic stars.

\section[]{The model of the flow structure in Z~And}
\label{sec:model}

According to the theoretical models \citep{Bis02,Mitsumoto05,Bis06} an accretion disk is formed around the compact object in the Z~And system when the wind of its giant has quiescent parameters. The radius of the disk is about 50$R_{\sun}$\, for a wind velocity of 20--25 \kms and thus the outer part of the disk is optically thin. The mass of the disk is estimated as product of one quarter of the mass-loss rate of the giant $\sim\!\!2\times10^{-7}M_{\sun}$ yr$^{-1}$ \citep{FC88}, and the typical time interval between the active phases---about 10 years. Based on the size and the mass of the disk of $\sim\!\!5 \times 10^{-7}M_{\sun}$\, we can assume that its inner region is optically thick in the quiescent state of the system.

To change the system from quiescent to active state a sufficiently large increase of the accretion rate is needed. Accretion of a considerable fraction of the disk's mass is required in order for an outburst to develop even in the combined model where the increased nuclear burning rate is taken into account \citep{Bis06}. Maximal increase of the accretion rate is possible in the framework of the mechanism proposed by \citet{Bis02} and \citet{Mitsumoto05}. According to this mechanism, even a small increase of the velocity of the wind of the donor is sufficient to change the accretion regime. During the transition from disk accretion to accretion from the flow, the disk is partially disrupted and the increased velocity of the wind causes falling of the material of the disk onto the accretor's surface. However, even in this case a considerable amount of mass (up to $50$--$80$ per cent) stays in the disk. Then massive accretion disk exists in the system during the active phase too.
\begin{figure}
	\includegraphics[width=0.47\textwidth]{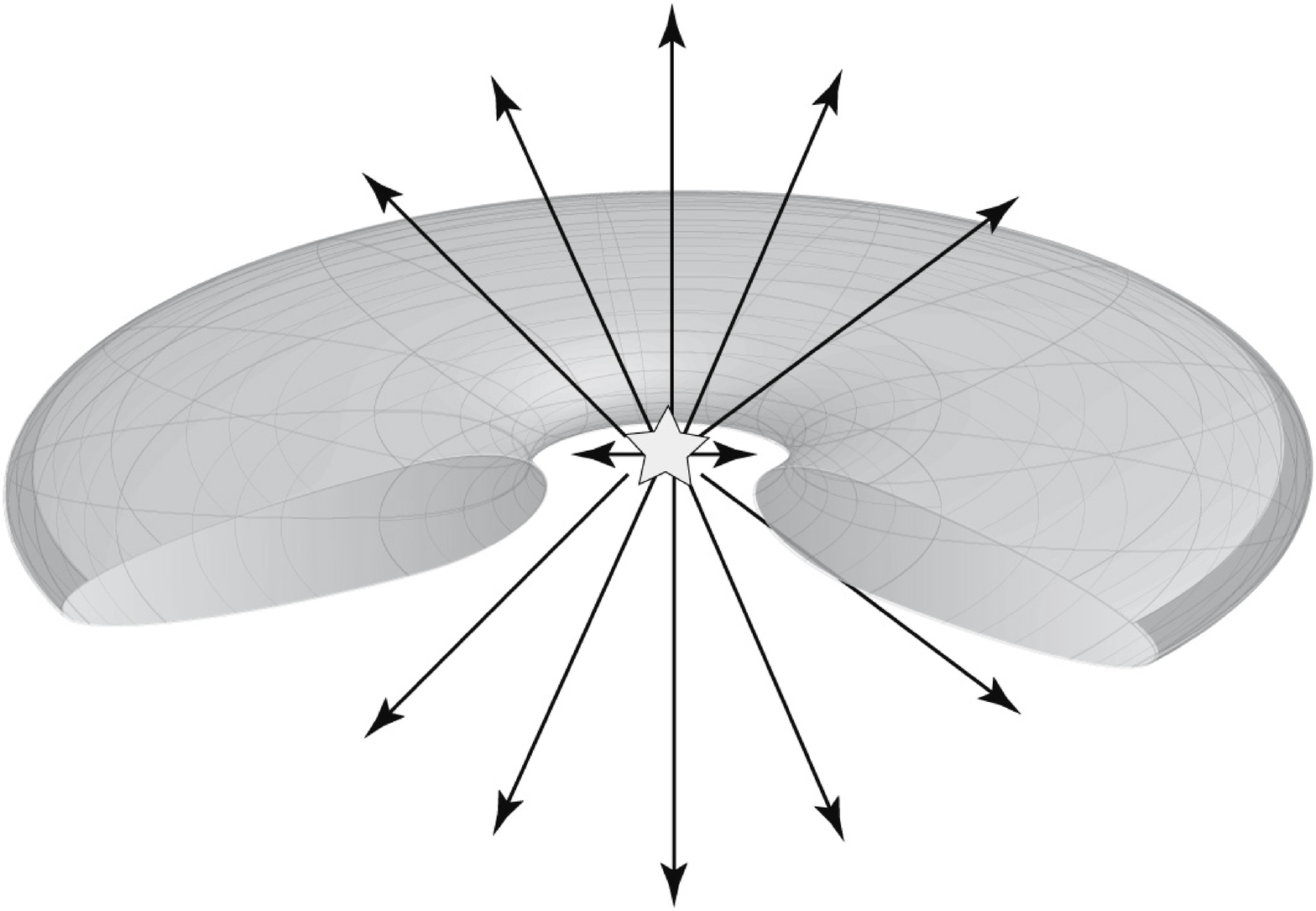}
	\includegraphics[width=0.47\textwidth]{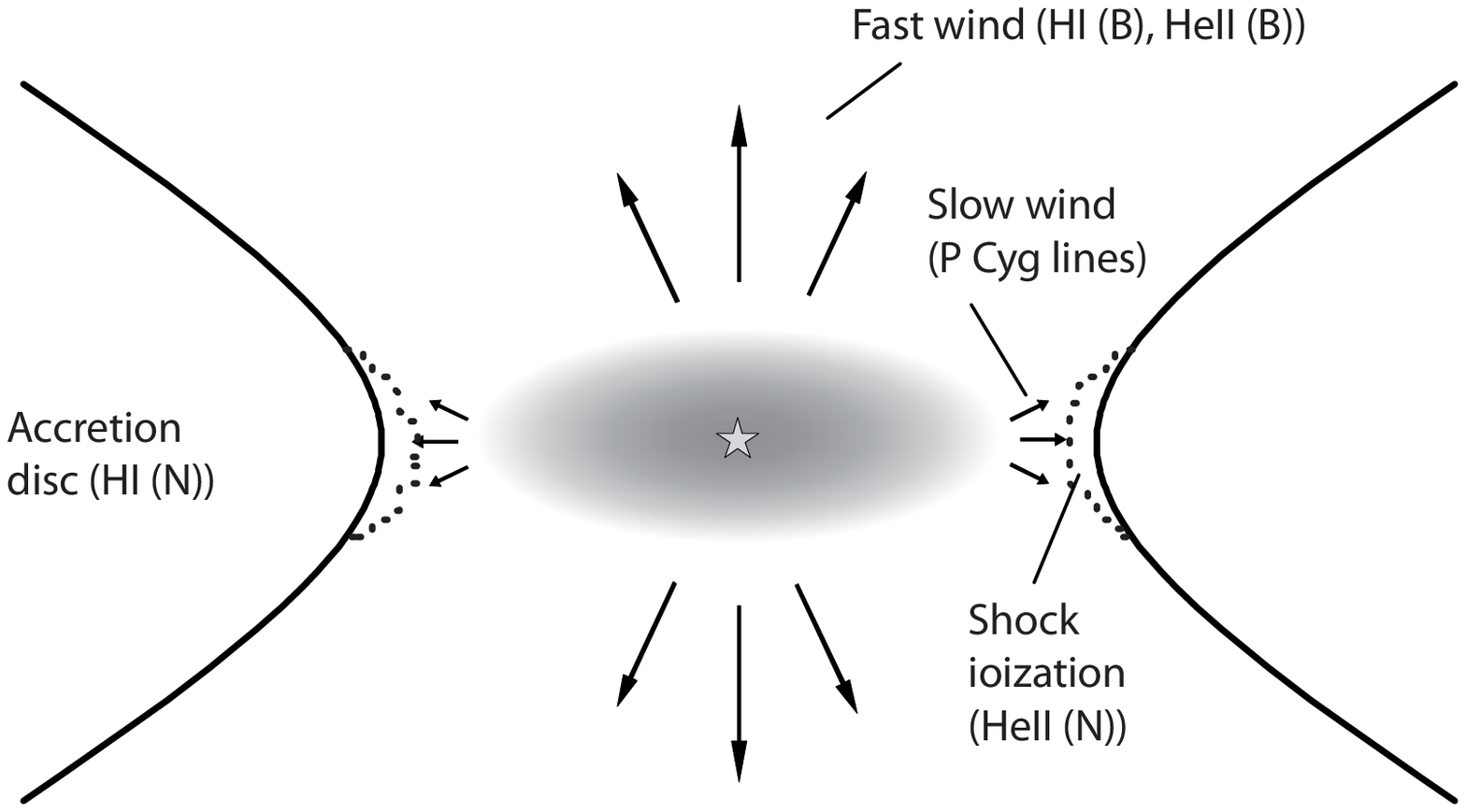}
\caption{Left panel: Schematic model of the region around the hot component during the first outburst.
Right panel: The same, but in the plane perpendicular to the orbital plane where the emission 	regions are shown. (From spectroscopy presented in \citet{TTB08}.)}
%	\label{Ha&Hg}
\end{figure}

During the outburst the high velocity wind of the hot compact component collides with the accretion disk, its velocity decreases near the orbital plane but does not change at higher stellar latitude. A consequence of this is that two different kinds of stellar wind are observed---P~Cyg wind with a low velocity and an optically thin high velocity wind. As a result of the collision the density of the wind close to the orbital plane increases and the level of the observed photosphere locates at a larger distance from the star. Thus, an optically thick disk-like shell forms, which occults the compact object. Since its effective temperature is lower it is responsible for the continuum energy redistribution and the increase of the optical flux of the star. The interaction of the wind with the disk is equivalent to interaction of two winds and thus can form shocked region of collisional ionization with temperature of up to $10^{6}$ K (Fig.~1) \citep{NW,Bis06}.

During the active phase the wind from the hot component ``strips'' the accretion disk carrying some part of its material in the circumbinary envelope. After each outburst some part of this material does not leave the potential well and after the cessation of the wind begins to accrete again. Because of the initial amount of angular momentum a disk-like envelope extending to larger distance from the orbital plane than the accretion disk itself forms (Fig.~2). The existence of a centrifugal barrier creates hollow cones around the axis of rotation with small opening angle \citep{I,BB}.
\begin{figure}[!htb]
	\includegraphics[width=0.47\textwidth]{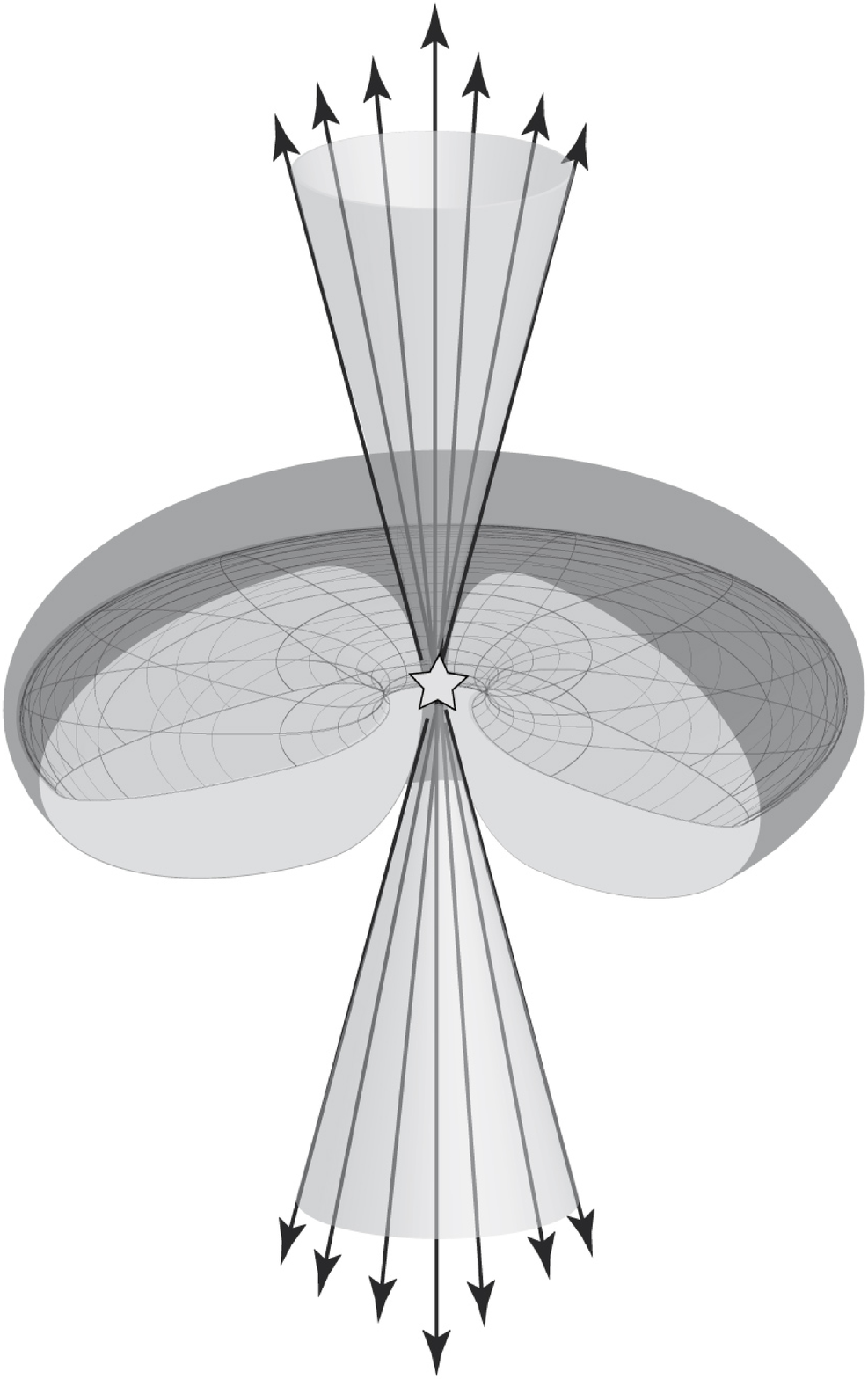}
	\includegraphics[width=0.47\textwidth]{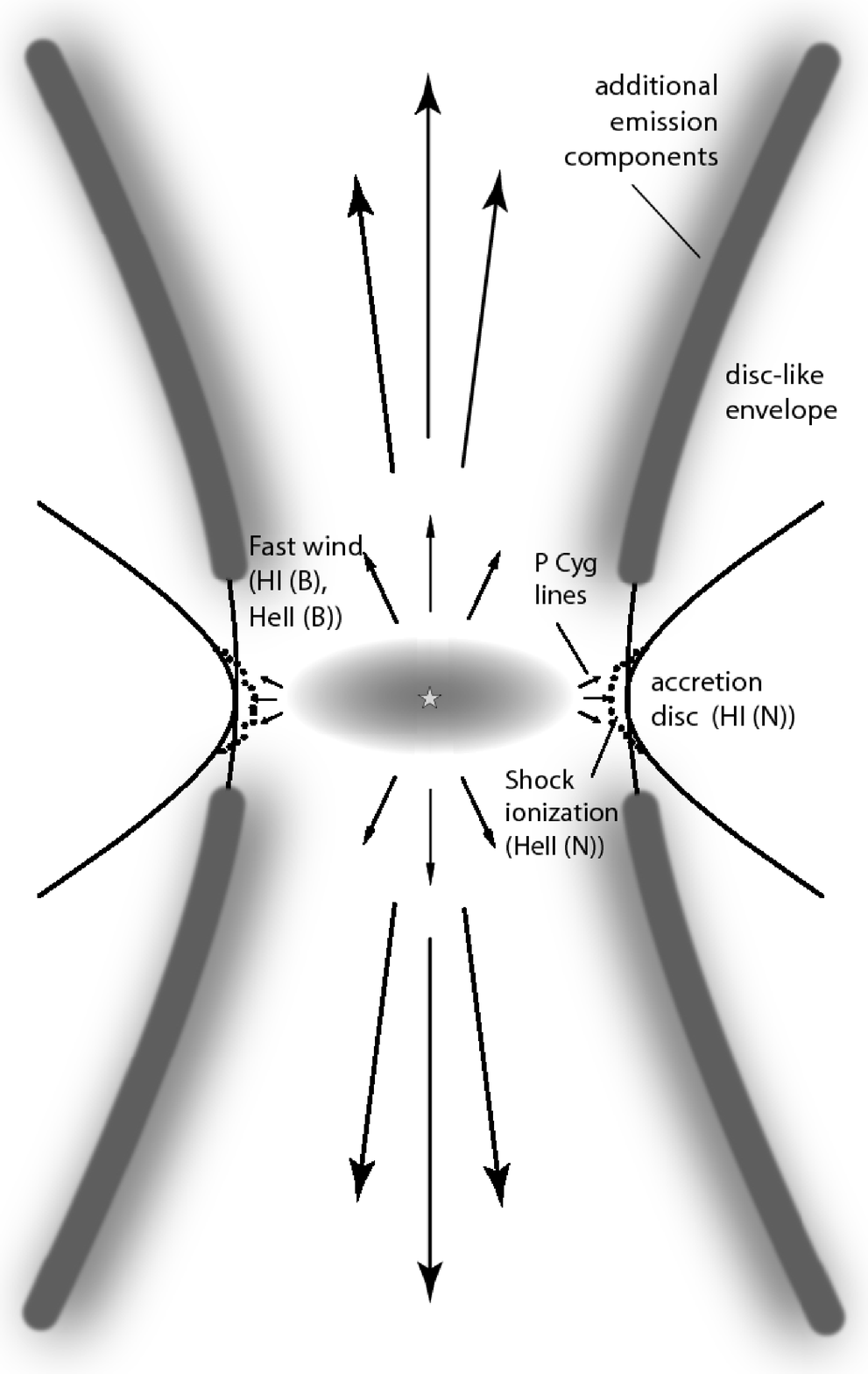}
 \caption{The same like in Fig.~1 but during recurrent strong outburst.}
\end{figure}

The disk-like envelope does not exist during the first outburst of the active phase. During the recurrent outbursts the extended disk-like envelope can collimate the wind which escapes the compact object predominantly via the two hollow cones (Fig.~2). The bipolar outflow can be observed as high-velocity satellite components situated on either side of the mean peak of the line. Their presence in the spectrum depends on the density of the disk-like envelope (i.e. on the system's activity during its previous phases) and on the wind intensity. They can appear only if the density of the disk-like envelope is sufficient to provide collimation and the mass-loss rate of the compact component is high enough. According to our model these spectral features can be observed during outbursts accompanied by loss of mass at high rate and preceded by similar strong outbursts.

According to the current theory the existence of bipolar collimated jets is supposed to be due to presence in the system of magnetized disk which transforms the potential energy of the accreting material into kinetic energy of the outflowing jets. This means that accretion must be realized to provide existence of the jets. In the spectrum of Z~And during its 2006 outburst, however, different indications of stellar wind presented along with satellite emission components. If we believe that ``traditional'' jets were observed, it is necessary to explain accretion onto the star at the time when a strong mass outflow from its surface was observed. On the other hand, the proposed model of collimated stellar wind makes it possible to explain all features of the observed spectrum without any important contradictions.

Another question related to the nature of the collimated outflow is that about the orbit inclination. The velocity of the radiatively accelerated wind cannot exceed $3000$ \kms. The highest observed outflow velocity during the 2006 outburst was $1500$ \kms. A preliminary analysis of our spectra obtained during late December 2009 showed velocities of $1700$--$1800$ \kms. Taking into account these velocities we obtain an upper limit of the orbit inclination of about $55$\degr. That is why the view about the nature of the outflow can shed some light on the question about the orbit inclination.

We suppose that the model suggested by us can be used to explain the behavior of the line spectrum of other classical symbiotic stars during their active phases too since the classical symbiotic stars have the next general characteristics:
\begin{enumerate}
	\item The mass transfer in the majority of the classical symbiotic stars is realized by means of the stellar wind of the giant. The theoretical computations of \citet{Bis02} and \citet{Mitsumoto05} show that for systems with parameters close to those of Z~And an accretion disk from wind accretion forms around the compact object. The accretion disk prevents the outflowing material during the active phase and outflow with two-component velocity regime forms.
	\item The systems with parameters close to Z~And have similar accretion rate of their compact object. The accretion rate determines the regime of hydrogen burning. The view that hydrogen burns in a steady state in the classical symbiotic stars is commonly accepted. When the accretion rate goes above the upper limit of the steady burning range the white dwarf expands which is observed as optical outburst with typical duration of one year. The first outburst opening the active phase can be followed by repeated outbursts. During the repeated outbursts an extended envelope surrounding the accretion disk exists in the system and it can be responsible for the collimation of the stellar wind.
\end{enumerate}

\section{Observational evidences in support of the model}
\label{sec:observ}		
\subsection{The first outburst}
\label{subsec:firstout}

According to our model a geometrically thin accretion disk exists in the system in the beginning of the first outburst and the inner region of this disk can be optically thick.

The presence of a moderate-velocity wind is indicated by the P~Cyg profiles of the triplet \mbox{He\,{\sc i}} $\lambda 4471$ and \mbox{He\,{\sc i}} $\lambda 4713$ lines during the light maximum. The two-component structure of their absorption showed mass outflow from the compact component with a mean velocity of $60$ \kms (Fig.~3) \citep{TTB08}.
\begin{figure}[!htb]
	\includegraphics[width=0.5\textwidth]{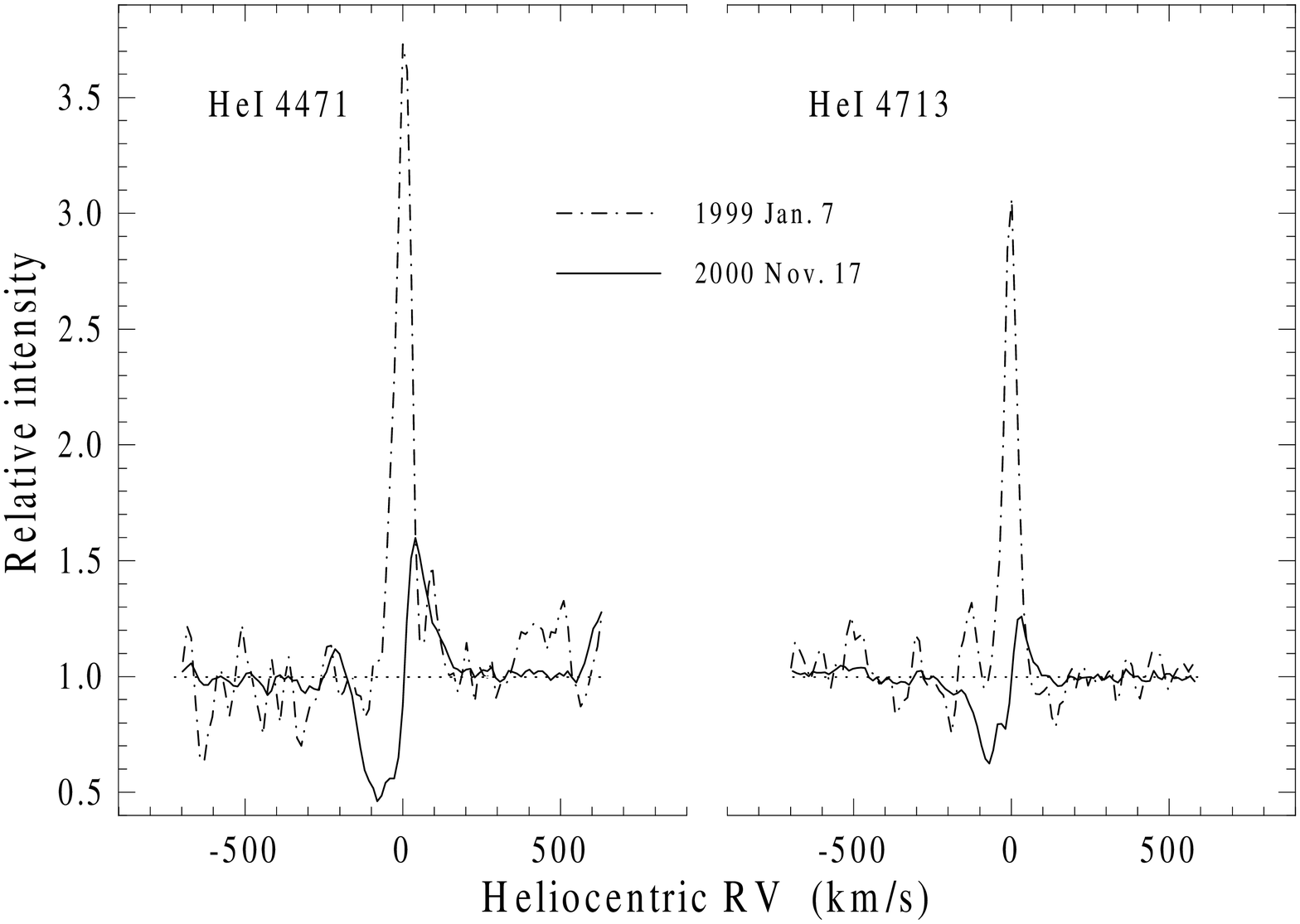}
%    \centering{\epsfig{file=fig3.ps, width=0.5\textwidth}}
 \caption{The profiles of the \mbox{He\,{\sc i}} triplet lines in quiescence (January 1999; dot-dashed line) and near the maximum of the first outburst (November 2000; solid line). (From spectroscopy presented in \citet{TTB08}.)}
\end{figure}
\begin{figure}[!htb]
	\includegraphics[width=0.5\textwidth]{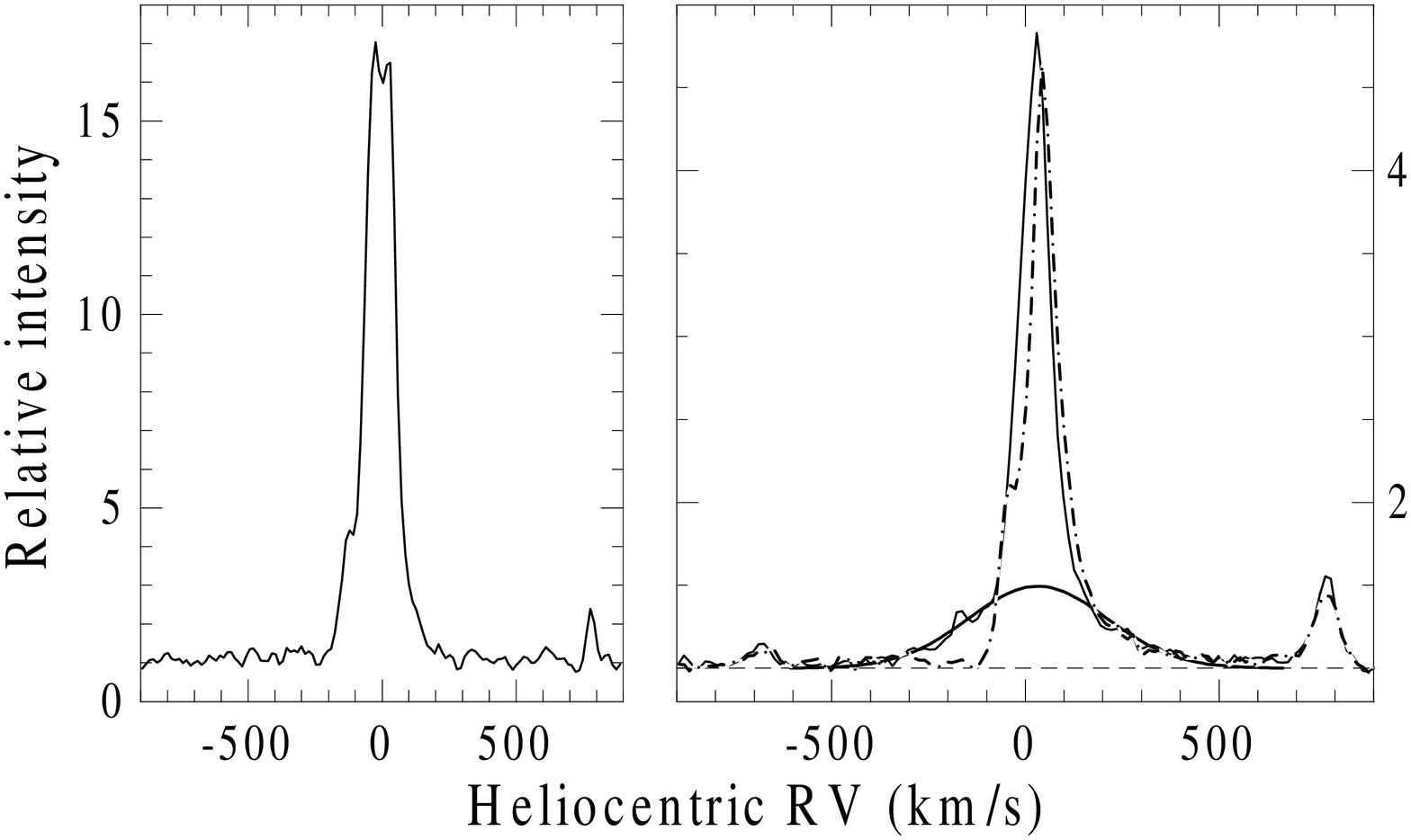}
%    \centering{\epsfig{file=fig4.ps, width=0.5\textwidth}}
 \caption{The profile of the H$_\gamma$ line in quiescence of Z And, on January 7, 1999 (left) and at various stages of the 2000 outburst (right). The dot-dashed line in the right panel corresponds to the spectrum on November 17, 2000, and the solid line---to the spectrum on December 5, 2000. The Gaussian fit of the broad component is marked with a thick line. The level of the local continuum is marked with a dashed line. (From spectroscopy presented in \citet{TTB08}.)}
\end{figure}

During the time of the increased light the H$_\gamma$ line had two emission components---a central narrow one and a broad, low-intensity component \citep{TTB08}. In November 2000 an absorption feature was probably present to the short-wavelength side of the narrow component, forming a dip in its emission profile (Fig.~4  right panel). The position of the dip corresponded to the red absorption component of the P Cyg lines of \mbox{He\,{\sc i}} (Fig.~5). The low velocity of the P~Cyg absorption and the dip in the H$_\gamma$ emission profile can be explained with supposition for decrease of the velocity of the wind of the compact component after its collision with the disk (Fig.~1(b)).

\begin{figure}[!htb]
	\includegraphics[width=0.45\textwidth]{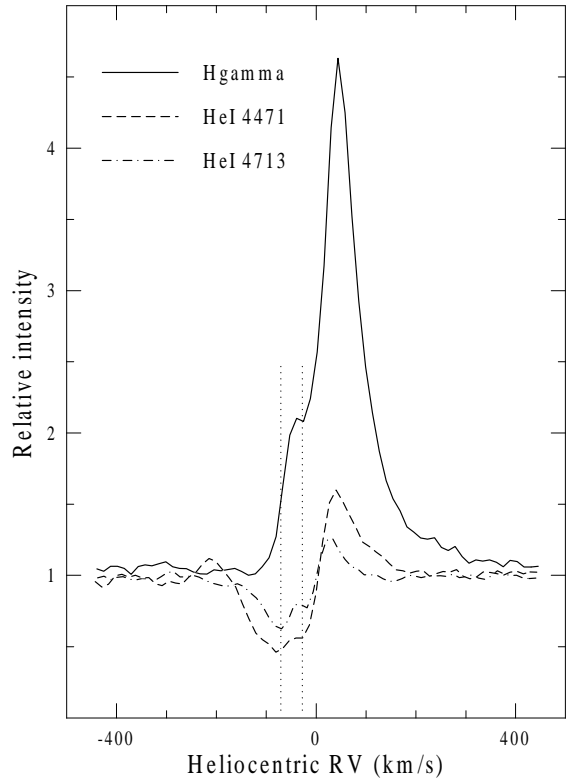}
%    \centering{\epsfig{file=fig5.ps, width=0.45\textwidth}}
 \caption{The profiles of the H$_\gamma$, \mbox{He\,{\sc i}} $\lambda$4471, and \mbox{He\,{\sc i}} $\lambda$4713 lines obtained on November 17, 2000. The positions of the two components of the \mbox{He\,{\sc i}} absorption are marked with vertical dotted lines. (From spectroscopy presented in \citet{TTB08}.)}
\end{figure}

The blue wing of the broad H$_\gamma$ component was not seen in November 2000 (Fig.~4, right panel). Assuming that the broad component (like the broad component of the \mbox{He\,{\sc ii}} $\lambda 4686$ line) is emitted by an optically thin high-velocity wind from the compact object, this can be explained with absorption in the stellar wind which is responsible for the P~Cyg profiles.

In the quiescent state an additional emission component was present on the short-wavelenghts side of the H$_\gamma$ line forming a shoulder in its profile. Thus the full width at the zero intensity (FWZI) of the line reached to about $500$ \kms\, (Fig.~4, left panel). The presence of such a component can be explained with emission by an accretion disc.
\begin{figure}[!htb]
	\includegraphics[width=0.5\textwidth]{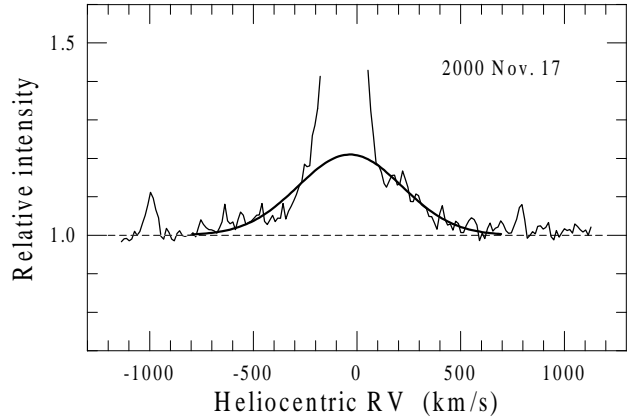}
%    \centering{\epsfig{file=fig6.ps, width=0.5\textwidth}}
 \caption{The wings of the \mbox{He\,{\sc ii}} $\lambda$4686 line during the 2000 outburst (November 17). The broad component was visible in this period. (From spectroscopy presented in \citet{TTB08}.)}
\end{figure}

The \mbox{He\,{\sc ii}} $\lambda 4686$ line of Z~And was two-component one consisting of a central narrow component and a broad component with a low intensity \citep{TTB08}. The broad component was very weak in quiescence and could not be measured accurately. It became stronger during the first outburst and towards the light maximum its FWZI exceeded $1000$ \kms\, (Fig.~6). We believe that during this period the broad component of the \mbox{He\,{\sc ii}} $\lambda 4686$ and H$_\gamma$ lines emitted in a region of stellar wind with a velocity of $500$ \kms\, traveling at higher stellar latitudes.

\subsection[]{The fourth outburst}
\label{subsec:fourthout}

The fourth outburst was the strongest one. The spectral data acquired during this outburst contain indications of all kinds of stellar wind, observed during the first one, and of bipolar collimated outflow in addition. We suppose that a disc-like envelope surrounding the accretion disc and collimating the outflowing material existed in the system at that time and the mass-loss rate was high enough to give rise to satellite line components.
\begin{figure}[!htb]
	\includegraphics[width=0.45\textwidth]{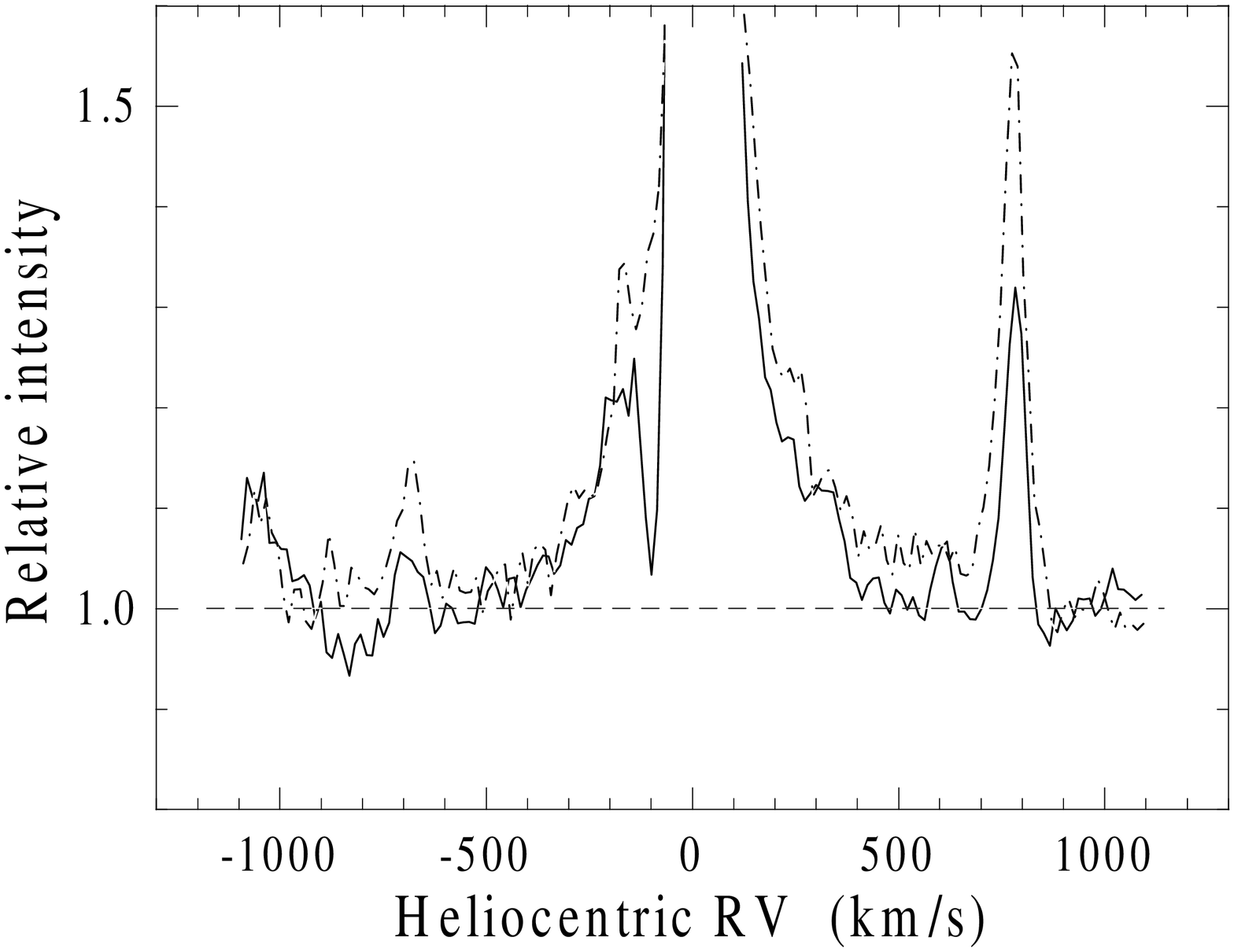}
%    \centering{\epsfig{file=fig7a.eps, width=0.45\textwidth}}
    \hspace{0.75 cm}
	\includegraphics[width=0.45\textwidth]{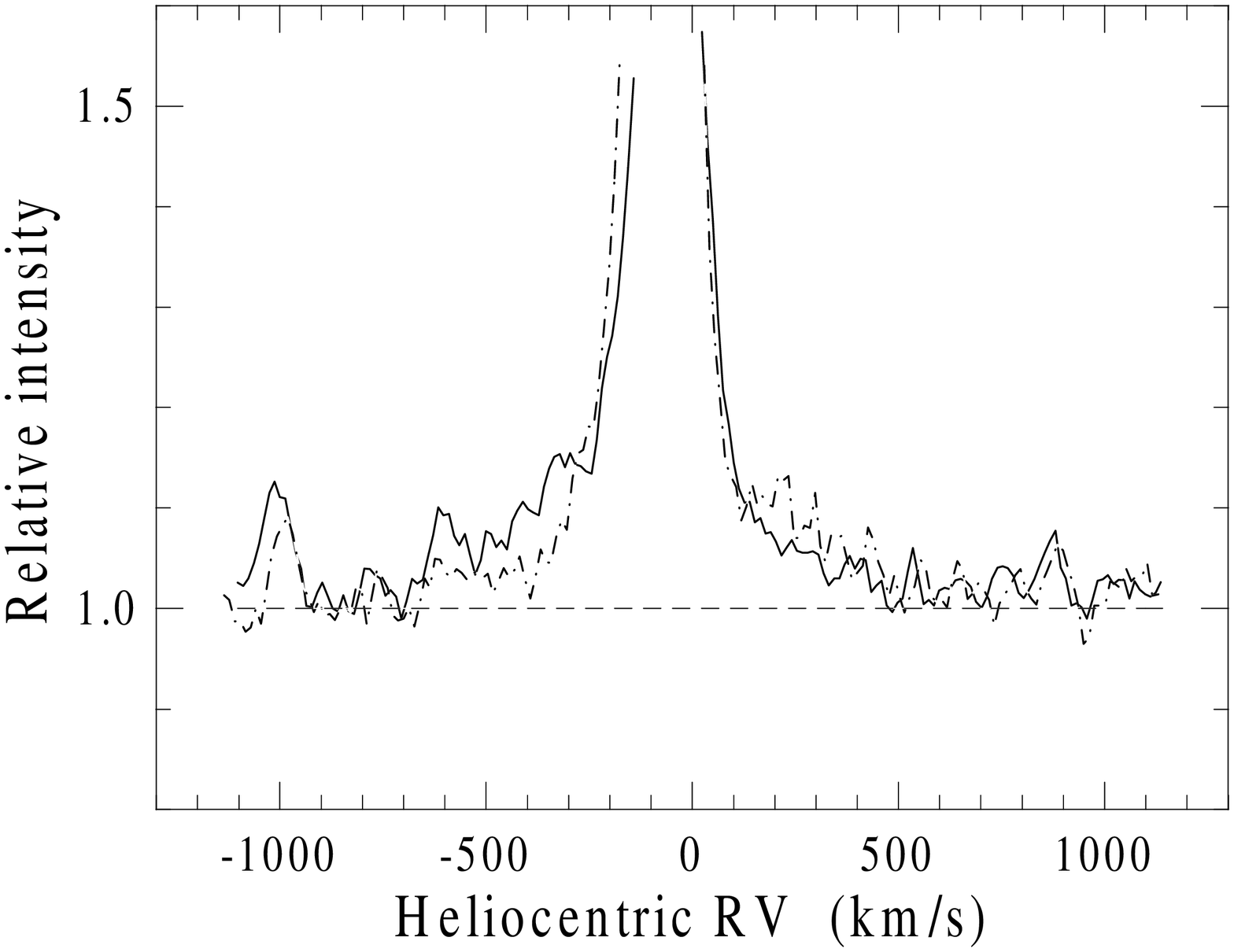}
%	    \centering{\epsfig{file=fig7b.eps, width=0.45\textwidth}}
 \caption{The broad component of the H$_\gamma$ (left panel) and \mbox{He\,{\sc ii}} $\lambda 4686$ (right panel) lines during the outbursts in 2000 (December 5; the dot-dashed line) and 2006 (October 4; the solid line). The level of the local continuum is marked with a dashed line.}
\end{figure}

The analysis of our observations of October 2006 with high resolution ($0.2$ \AA\, per pixel, as for the first and second outbursts) in the H$_\gamma$ and \mbox{He\,{\sc ii}} $\lambda 4686$ region of the spectrum of Z~And reveals the presence of emission up to $500$ \kms\, from the centers of these lines (Fig.~7). \citet{Sk09} obtained the effective photospheric radius of the outbursting component of $(12 \pm 4)R_{\sun}$ at the time of the light maximum for a distance of $1.5$ kpc. The minimal radius for the distance of $1.12$ kpc used by us \citep{TTB08,TTB10} is $6R_{\sun}$. If the broad emission originated in a Keplerian accretion disk at a distance equal to the radius of the outbursting component, its velocity should be not greater than $140$ \kms, in disagreement with the observations. Like in the case of the first outburst it is natural to suppose that the observed broad emission is formed in the high-velocity wind at higher stellar latitudes (Fig.~2). A dip is observed in the blue part of the broad H$_\gamma$ emission at a velocity of about $100$ \kms.
\begin{figure}[!htb]
	\includegraphics[width=0.5\textwidth]{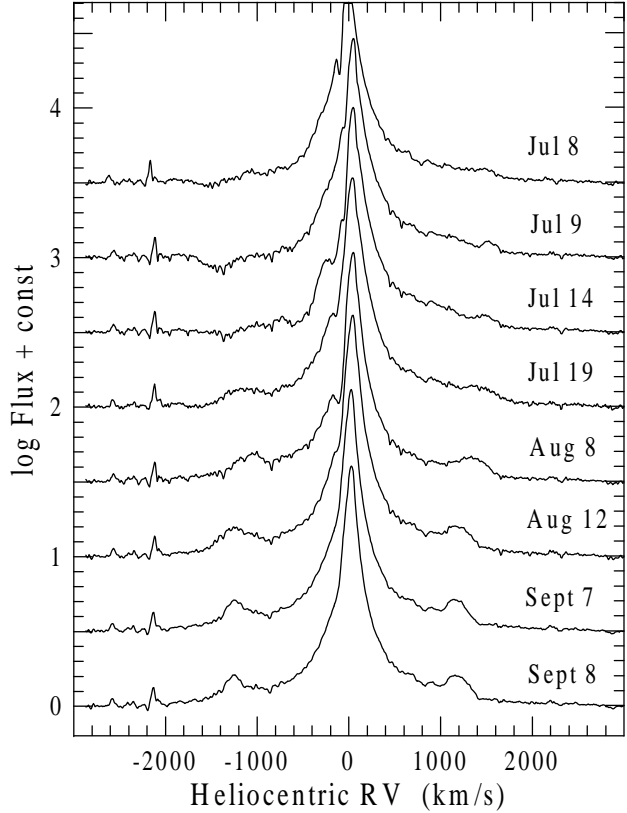}
%    \centering{\epsfig{file=fig8.ps, width=0.5\textwidth}}
 \caption{Evolution of the H$_\alpha$ line during the 2006 outburst. (From spectroscopy presented in \citet{TTB07}.)}
\end{figure}

In our spectra \citep{TTB07}, the H$_\alpha$ line possessed a strong central component, a broad component (wings) and additional absorption and emission features to either side of the central component (Fig.~8). The central component was single peaked having a shoulder on its short-wavelength side, which was not seen only in the spectra taken in September. A small peak shortward from the central component was seen in the spectrum obtained on August 8, 2006, and the dip before the peak had a velocity about $100$ \kms, in agreement with the results of the analysis of the H$_\gamma$ and \mbox{He\,{\sc i}} profiles. Satellite emission components on either side of the central peak appeared after mid-July (Fig.~8). They are clear signs of the presence of a bipolar collimated wind. According to our model the emitting region is near the cone surfaces (see Fig.~2), where the density of the emerging collimated outflow is the highest.
\begin{figure}[!htb]
	\includegraphics[width=0.7\textwidth]{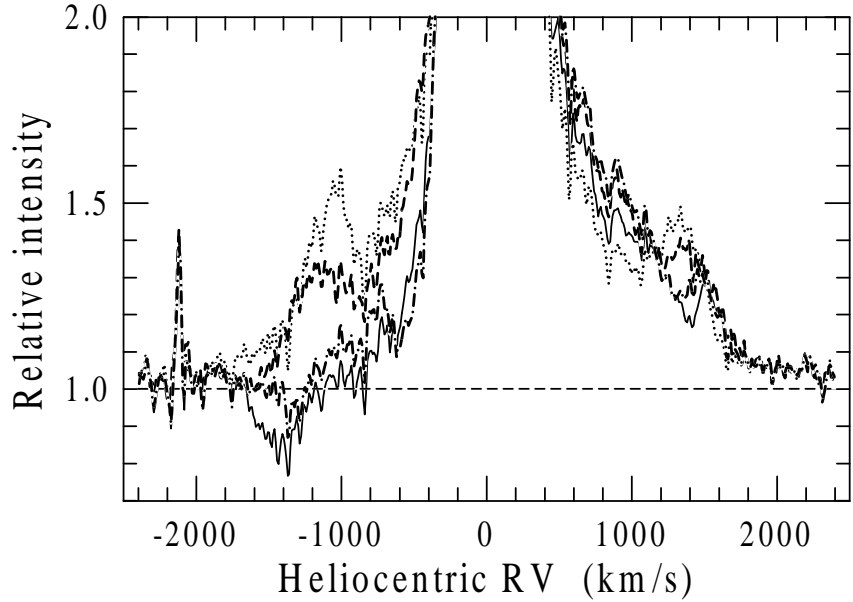}
%    \centering{\epsfig{file=fig9.ps, width=0.7\textwidth}}
 \caption{The variation of the H$_\alpha$ line: transition from absorption feature to emission one. The solid line corresponds to the spectrum taken on July 9, 2006; the dot-dashed line to July 14, 2006; the dotted line to August 8, 2006; and the dashed line to September 7, 2006. The level of the local continuum is marked with a dashed line. (From spectroscopy presented in \citet{TTB07}.)}
\end{figure}

The first spectra during this outburst taken in July 2006 (Figs.~8 and 9) show a sharp absorption line at a velocity of 1400 \kms\, on the short-wavelength side of the central component and only a weak irregularly shaped emission component corresponding to a velocity of about $1500$ \kms\, on its long-wavelength side. The blue-shifted absorption probably indicates the flow which is projected onto the stellar disk. The evolution of the spectrum in Figs.~8 and 9 shows that the blue-shifted absorption component disappears and an emission appears. Two emission components on either side of the central peak were formed in July. The disappearance
of the blue-shifted absorption component and the appearance of emission are probably due to a decrease of the mass-loss rate and/or an increase of the number of emitting atoms in the region of the wind that does not project onto the stellar disk.

During all the outbursts considered by us the H$_\alpha$ line possessed broad wings extending to velocities
of at least $\pm 2000$ \kms\, from its center \citep{TTB07,TTB08,TTB10}.
\citet{Sk06} concluded that the H$_\alpha$ wings during the outburst state were formed in a high-velocity stellar wind from the compact component. However, as it was noted in \citet{TTB10} it is not ruled out that the high-velocity broadening of the H$_\alpha$ line was due to radiative damping and the stellar wind contributed to the emission at lower velocities.

\section[]{Mass-loss rate}
\label{sec:mass-loss}
	
The mass-loss rate of the compact object during the existence of the collimated wind was calculated from the energy flux of the broad H$_\gamma$ component, observed on October 31, 2006 which was analyzed by fitting with a Gaussian function (Fig. 10). This broad component was not symmetric since its blue wing was absorbed by the wind outflow responsible for the P~Cyg absorption components like during the first outburst. Having in mind this absorption we consider the wind velocity obtained by us as one lower limit. The equivalent width of the broad component was obtained with an error of 10 per cent depending mainly on the level of the local continuum. The real error however, is greater since the broad emission is not clearly visible because of blending with the central narrow component and its profile is actually not Gaussian. The energy flux of the broad component was obtained using its equivalent width and the continuum flux at the position of the line H$_\gamma$. This continuum flux was calculated with linear extrapolation of the B and V fluxes from the photometric data of \citep{Sk07} taken very close to the time of our observations. The uncertainty of the continuum flux is not more than $10$ per cent. The line flux was corrected for an interstellar extinction of $E(B - V) = 0.30$ according to the approach used in \citet{CCM89}.
\begin{figure}[!htb]
	\includegraphics[width=0.5\textwidth]{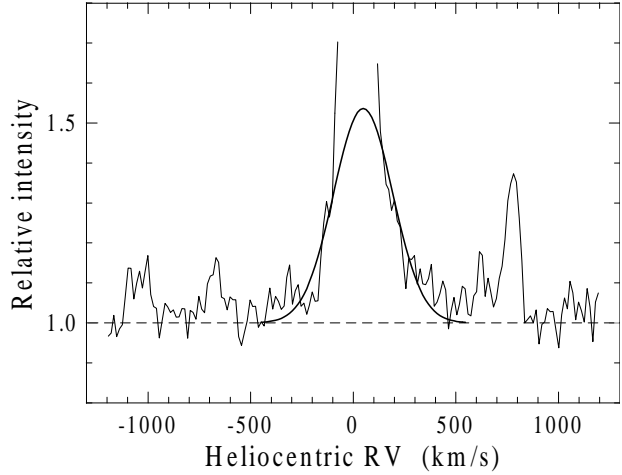}
%    \centering{\epsfig{file=fig10.eps, width=0.5\textwidth}}
 \caption{The broad component of the H$_\gamma$ line on October 31, 2006. The level of the local continuum is marked with a dashed line.}
\end{figure}

The mass-loss rate was calculated supposing that the medium of the wind is optically thin and that the outflow has a spherical symmetry and a constant velocity. The particle density in the wind is expressed via the continuity equation. In our calculations, we adopted a value of the electron temperature in the wind of $30\,000$ K like during the first outburst \citep{TTB08}. We used a parameter $\mu = 1.4$ \citep{NV87} determining the mean molecular weight $\mu m_{\rm H}$ in the wind and a helium abundance of 0.1 \citep{VN94}. We adopted a distance to the system $d = 1.12$ kpc \citep{FC88,FC95} as it was used in the paper of \citet{TTB08} to compare the results more easily. It is supposed that the line is emitted by a spherical layer and the radii of integration must be estimated. We assumed optically thin medium and the inner radius in this case is thought to be the photospheric radius. The photospheric radius was estimated from the bolometric luminosity and the effective temperature of the outbursting compact object at the time of our observation. We used a bolometric luminosity of $10^{4}L_{\sun}$ \citep{Sk09} and Zanstra temperature of $90\,000$ K \citep{BL} and obtained a photospheric radius of $0.31(d/1.12$ kpc)$R_{\sun}$. The outer radius of integration is equal to infinity. We used a recombination coefficients for case B \citep{SH} corresponding to temperature of $30\,000$ K and the density at the level of the photosphere. Having a line flux of $7.83\times10^{-12}$ erg\,cm$^{-2}$\,s$^{-1}$ and a velocity of the wind of $370$ km\,s$^{-1}$ we obtained a mass-loss rate of $1.33\times10^{-7} (d/1.12$ kpc)$^{3/2}M_{\sun}$ yr$^{-1}$.

The total mass of the two parts (jets) of the collimated wind where the emitting and the non emitting masses are included, is equal to the mass, ejected by the system during the time of existence of the jets (July--December 2006). The ejected mass is $m = \dot M \Delta t$ , where $\dot M$ is the mass-loss rate of the outbursting compact object and $\Delta t$---the time of existence of the jets. Then the total mass of the two jets is obtained to be about $0.7\times10^{-7} (d/1.12$ kpc)$^{3/2}M_{\sun}$.

\section[]{Conclusions}
\label{sec:concl}

The comparison of the results of our observations with the results of the gas-dynamical modeling provides possibility to propose a model to explain the behavior of the Z~And system during its whole 2000--2010 active phase as well as the differences in its observed properties during different outbursts. The basic points of this model are the following ones:
\begin{itemize}
\renewcommand{\labelitemi}{$\bullet$}
	\item An accretion disk exists in the system.

	\item During the outburst the gas leaving the surface of the compact component at high velocity collides with the accretion disk. As a result its velocity decreases near the orbital plane, but remains unchanged at higher latitudes. Hence, two wind components with different velocities are observed.

	\item During the outburst the wind from the compact component partially disrupts the disk. Some portion of its mass does not leave the potential well, after the wind ceases it begins to accrete again. Because of the initial amount of angular momentum a disk-like envelope extending to larger distance from the orbital plane forms.

	\item The existence of a centrifugal barrier creates hollow cones around the axis of rotation which lead to appearance of collimated outflow during the recurrent outbursts. This outflow can be observed as high-velocity components on either side of the main peak of the line.

	\item Collimated outflow will appear when the density of the disk-like envelope is high enough to provide collimation and the mass-loss rate of the compact component is also high. High-velocity line components thus can be observed during outbursts when the loss of mass is at high rate, preceded by a similarly strong outburst.
\end{itemize}

The model suggested provides possibility to explain all the spectral features of Z~And observed during active stage. The general character of this model permits us to suppose that similar scenario is possible for other classical symbiotic stars too.

\begin{theacknowledgments}

This study was partially supported by the Russian and Bulgarian Academies of Sciences through a collaborative program in basic space research, the Russian Foundation for Basic Research (Project Code 09-02-00064), the Federal Target Program on ``Human Resources for Science, Research, Education and Innovation in Russia'' for 2009--2013, and the Bulgarian Scientific Research Fund (Grant
DO~02-85).

\end{theacknowledgments}


\begin{thebibliography}{25}

%\bibliography{}

\bibitem[M\"urset \& Schmid(1999)]{MS}
	U.~M\"urset and H.~M.~Schmid, \emph{Astron.\ Astrophys.\ Suppl.}\ \textbf{137}, 473--493 (1999).

\bibitem[Sokoloski et al.(2006)]{Sok06}
	J.~L.~Sokoloski, S.~J.~Kenyon, B.~R.~Espey, Charles D.~Keyes, S.~R.~McCandliss, A.~K.~H.~Kong, J.~P.~Aufdenberg,
    A.~V.~Filippenko, W.~Li, C.~Brocksopp, Christian~R.~Kaiser, P.~A.~Charles, M.~P.~Rupen, and R.~P.~S.~Stone, \emph{Astrophys.\ J.}\ \textbf{636}, 1002--1019 (2006).

\bibitem[Swings \& Struve(1941)]{SS}
	P.~Swings and O.~Struve, \emph{Astrophys.\ J.}\ \textbf{93}, 356--368 (1941).

\bibitem[Boyarchuk(1967)]{Boyarchuk}
	A.~A.~Boyarchuk, \emph{Soviet Astron.}\ \textbf{11}, 818--827 (1967).

\bibitem[Fernandez-Castro et al.(1995)]{FC95}
	T.~Fernandez-Castro, R.~Gonzalez-Riestra, A.~Cassatella,
	A.~Taylor, and E.~R.~Seaquist, \emph{Astrophys.\ J.}\ \textbf{442}, 366--380 (1995).

\bibitem[Bisikalo et al.(2006)]{Bis06}
	D.~V.~Bisikalo, A.~A.~Boyarchuk, E.~Yu.~Kilpio, N.~A.~Tomov, and M.~T.~Tomova, \emph{Astron.\ Rep.}\ \textbf{50}, 722--732 (2006).

\bibitem[Skopal et al.(2006)]{Sk06}
	A.~Skopal, A.~A.~Vittone, L.~Errico, M.~Otsuka, S.~Tamura, M.~Wolf, and V.~G.~Elkin, \emph{Astron.\ Astrophys.}\ \textbf{453}, 279--293 (2006).

\bibitem[Tomov et al.(2008)]{TTB08}
	N.~A.~Tomov, M.~T.~Tomova, and D.~V.~Bisikalo, \emph{Mon.\ Not.\ R.\ Astron.\ Soc.}\ \textbf{389}, 829--838 (2008).

\bibitem[Skopal et al.(2009)]{Sk09}
	A.~Skopal, T.~Pribulla, J.~Budaj,  A.~A.~Vittone, L.~Errico, M.~Wolf, M.~Otsuka, M.~Chrastina, and	
	Z.~Mikulasek, \emph{Astrophys.\ J.}\ \textbf{690}, 1222--1235 (2009).

\bibitem[Tomov et al.(2010)]{TTB10}
	N.~A.~Tomov, M.~T.~Tomova, and D.~V.~Bisikalo, \emph{Astron.\ Rep.}\ \textbf{54}, 528--536 (2010).

\bibitem[Skopal et al.(2000)]{Sk00}
	A.~Skopal, D.~Chochol, T.~Pribulla, and M.~Vanko, \emph{Inform.\ Bull.\ Var.\ Stars\/} No.~5005 (2000).

\bibitem[Tomov et al.(2007)]{TTB07}
	N.~A.~Tomov, M.~T.~Tomova, and D.~V.~Bisikalo, \emph{Mon.\ Not.\ R.\ Astron.\ Soc.}\ \textbf{376}, L16--L19 (2007).

\bibitem[Burmeister \& Leedjarv(2007)]{BL}
	M.~Burmeister and L.~Leedjarv, \emph{Astron.\ Astrophys.}\ \textbf{461}, 5L--8L (2007).

\bibitem[Bisikalo et al.(2002)]{Bis02}
	D.~V.~Bisikalo, A.~A.~Boyarchuk, E.~Yu.~Kilpio, and O.~A.~Kuznetsov, \emph{Astron.\ Rep.}\ \textbf{46}, 1022--1029 (2002).

\bibitem[Mitsumoto et al.(2005)]{Mitsumoto05}
     M.~Mitsumoto, B.~Jahanara, T.~Matsuda, K.~Oka, D.~V.~Bisikalo, E.~Yu.~Kilpio, H.~M.~J.~Boffin, A.~A.~Boyarchuk, and
     O.~A.~Kuznetsov, \emph{Astron.\ Rep.}\ \textbf{49}, 884--891 (2005).

\bibitem[Fernandez-Castro et al.(1988)]{FC88}
	 T.~Fernandez-Castro, A.~Cassatella, A.~Gimenez, and R.~Viotti,
 \emph{Astrophys.\ J.}\ \textbf{324}, 1016--1025 (1988).

\bibitem[Nussbaumer \& Walder(1993)]{NW}
	H.~Nussbaumer and R.~Walder, \emph{Astron.\ Astrophys.}\ \textbf{278}, 209--225 (1993).

\bibitem[Icke(1981)]{I}
	V.~Icke, \emph{Astrophys.\ J.}\ \textbf{247}, 152--157 (1981).

\bibitem[Blandford \& Begelman(2004)]{BB}
	R.~D.~Blandford and M.~C.~Begelman, \emph{Mon.\ Not.\ R.\ Astron.\ Soc.}\ \textbf{349}, 68--86 (2004).

\bibitem[Skopal et al.(2007)]{Sk07}
	A.~Skopal, M.~Vanko, T.~Pribulla, D.~Chochol, E.~Semkov, M.~Wolf, and A.~Jones,
	\emph{Astron.\ Nachr.}\ \textbf{328}, 909--916 (2007).

\bibitem[Cardelli et al.(1989)]{CCM89}
	J.~A.~Cardelli, G.~C.~Clayton, and J.~S.~Mathis, \emph{Astrophys.\ J.}\ \textbf{345}, 245--256 (1989).

\bibitem[Nussbaumer \& Vogel(1987)]{NV87}
	H.~Nussbaumer and M.~Vogel, \emph{Astron.\ Astrophys.}\ \textbf{182}, 51--62  (1987).

\bibitem[Vogel \& Nussbaumer(1994)]{VN94}
	M.~Vogel and H.~Nussbaumer, \emph{Astron.\ Astrophys.}\ \textbf{284}, 145--155 (1994).

\bibitem[Storey \& Hummer(1995)]{SH}
	P.~J.~Storey and D.~G.~Hummer, \emph{Mon.\ Not.\ R.\ Astron.\ Soc.}\ \textbf{272}, 41--48 (1995).

\end{thebibliography}
\end{document}